\documentclass[preprintnumbers, prd, onecolumn, showpacs,floatfix,preprintnumbers,
superscriptaddress, nofootinbib]{revtex4}
\usepackage{graphicx}
\usepackage{epsfig}
\usepackage{bm}
\usepackage{amssymb}
\usepackage{float}
\usepackage{amsmath}
\usepackage{subfigure}
\usepackage{dcolumn}
\usepackage{cancel}
\usepackage[colorlinks]{hyperref}
\usepackage[usenames,dvipsnames]{color}
\hypersetup{
     breaklinks=true,
    pdfstartview={FitH},    
    colorlinks=true,       
    linkcolor=blue,          
    citecolor=red,        
    filecolor=magenta,      
    urlcolor=blue,           
    anchorcolor=green,      
    linktocpage=true
}

\def\doi{http://doi.org}

\begin{document}
\title{The Logotropic Dark Fluid: Observational and Thermodynamic Constraints}
\author{Abdulla Al Mamon}
\email{abdulla.physics@gmail.com}
\affiliation{Department of Physics, Vivekananda Satavarshiki Mahavidyalaya (affiliated
to the Vidyasagar University), Manikpara-721513, West Bengal, India}
\author{Subhajit Saha}
\email{subhajit1729@gmail.com}
\affiliation{Department of Mathematics,
Panihati Mahavidyalaya, Kolkata-700110, West Bengal, India}
\newcommand{\be}{\begin{equation}}
\newcommand{\ee}{\end{equation}}
\newcommand{\bea}{\begin{eqnarray}}
\newcommand{\eea}{\end{eqnarray}}
\newcommand{\bc}{\begin{center}}
\newcommand{\ec}{\end{center}}
\begin{abstract}
We have considered a spatially flat, homogeneous and isotropic FLRW Universe filled with a single fluid, known as logotropic dark fluid (LDF), whose pressure evolves through a logarithmic equation of state. We use the recent Pantheon SNIa and cosmic chronometer datasets to constrain the parameters of this model, the present fraction of dark matter $\Omega_{m0}$ and the Hubble constant $H_0$. We find that the mean values of these parameters are $\Omega_{m0}=0.288\pm 0.012$ and
$H_{0}=69.652\pm 1.698~{\rm km/s/Mpc}$ at the $1\sigma$ CL.
We also find that the LDF
model shows a smooth transition from the deceleration phase to acceleration phase of the universe in the recent past. We notice that the redshift of this transition $z_{t}=0.706\pm 0.048$ ($1\sigma$ error) and is well consistent with the present observations. Interestingly, we find that the Universe will settle down to a $\Lambda$CDM model in future and there will not be any future singularity in the LDF model. Furthermore, we notice that there is no significant difference between the LDF and $\Lambda$CDM models at the present epoch, but the difference (at the percent level) between these models is found as the redshift increases. We have also studied the generalized second law of thermodynamics at the dynamical apparent horizon for the LDF model
with the Bekenstein and Viaggiu entropies.  
\end{abstract} 
\maketitle
\section{Introduction}\label{secintro}
Many cosmological observations \cite{acc1,acc2,acc3,acc4,acc5,acc6,acc7,acc8,acc9,acc10} have supported that the expansion of the current Universe is accelerating and the alleged acceleration is rather a recent phenomenon. In this context, the most accepted idea is that an exotic component of the matter sector with long range anti-gravity properties, dubbed as {\it dark energy}, is responsible for this acceleration mechanism. However, the true nature of dark energy (DE) and dark matter (DM) is still unknown and a plethora of theoretical models has been introduced to account for the observation of cosmic acceleration (for review, one can look into Refs. \cite{de1,de2,de3}). In the context of DE, the concordance Lambda-Cold-Dark-Matter ($\Lambda$CDM) model is the simplest model and is consistent with most of the observations \cite{acc10}. However, this model suffers from the {\it fine tuning} and the cosmological {\it coincidence} problems \cite{ccp1,ccp2}. Till now, we do not have a concrete cosmological model that can provide a satisfactory solution to all the problems.\\
\par Recently, Logotropic Dark Fluid (LDF), a robust and natural candidate for unifying DE and DM, has gained immense interest in the literature \cite{lmc1,lmc2,lmc3,lmc4,lmc5,lmc6,lmc7}. An important advantage lies on the fact that it is a consequence of the first principle of thermodynamics. The LDF, proposed by P.H. Chavanis \cite{lmc1,lmc2}, is an attempt towards unification of DM and DE. It belongs to the class of modified matter models in an otherwise flat, homogeneous and isotropic FLRW Universe. Capozziello et al. \cite{lmc7} recently introduced a new class of barotropic factor for matter based on the properties of isotropic deformations of crystalline solids. They dubbed their approach as Anton-Schmidt’s equation of state which gives a small, yet non-vanishing pressure term for matter. This means that the corresponding pressure is proportional to the logarithm of the volume of the universe. Their model is an extension of the LDF and it contains two free parameters, $B$ and $n$. A year later, Boshkayev et al. \cite{lmc6} studied the generalization of logotropic models. They showed that these models form a subset within the picture of an Anton-Schmidt fluid. In doing so, they have also been able to recover the modified Chaplygin gas under certain conditions. In a pioneering work, Chavanis and Kumar \cite{lmc3} performed a detailed comparison between the LDF model and the $\Lambda$CDM model at the cosmological (large) scales. Using the observational data
from Planck 2015+Lensing+BAO+JLA(SNIa)+HST, they have found that the best fit values of $\Omega_{m0}$ and $H_{0}$ are $\Omega_{m0}=0.3014$ and $H_{0}=68.30~{\rm km~s^{-1}~Mpc^{-1}}$ for the LDF model, and $\Omega_{m0}=0.3049$ and $H_{0}=68.02~{\rm km~s^{-1}~Mpc^{-1}}$ for the $\Lambda$CDM model. It is worth noting that the Logotropic model is almost indistinguishable from the $\Lambda$CDM model for a substantial part of the late-time evolution of the Universe right up to the present time. However, the difference between the two models will be reflected at some point in the future (about 25 Gyrs  from now) when the LDF behaves as a phantom fluid, while the $\Lambda$CDM model will enter in a de Sitter phase \cite{lmc3}. Additionally, the advantages of considering a LDF is three fold \cite{lmc3} --- (a) The speed of sound and the Jeans length are both non-zero in a Logotropic model which might alleviate the cusp problem and the missing satellite problem of the $\Lambda$CDM model, (b) Such a model is consistent with the empirical Burkert profile of galaxy rotation curves \cite{Burkert1} which are characterstic of most observed DM halos. This is not the case with the $\Lambda$CDM model \cite{Navarro1}, and (c) The universality of the surface density of DM halos \cite{Donato1}, the universality of the mass of dwarf spheroidal galaxies \cite{Strigari1}, and the Tully-Fisher relation \cite{Tully1} are explained neatly by the Logotropic model, as confirmed by analysis of observational data \cite{lmc1,lmc2}. These remarkable features have placed the LDF in a unique spot amongst other unified models \cite{udem1,udem2,udem3,udem4,udem5} which attempt to unify DM and DE. \\
\par Motivated by the above facts, in this paper, we consider that the Universe is made of a single dark fluid described by a logotropic equation of state. The cosmological aspects of this model has already been studied in Refs. \cite{lmc1,lmc2}. In the present work, using the latest Pantheon SNIa and cosmic chronometer datasets, we try to constrain the model parameters to study the different properties of this model extensively. By considering the Universe as a thermodynamical system, we also study the thermodynamics of the model at the dynamical apparent horizon, particularly, the {\it generalized second law} (GSL) of thermodynamics. For this purpose, we consider two different entropies, viz, Bekenstein and Viaggiu entropies \cite{Bekenstein1,Viaggiu1,Viaggiu2}. With this thermodynamic analysis, we also try to constrain the dimensionless logotropic temperature $B$ whose value has been obtained from the surface density of DM halos by Chavanis \cite{lmc1}. In this regard, it is worthwhile to mention that Tian and Booth \cite{Tian1} have pointed out several controversial aspects of gravitational thermodynamics and also made an attempt to address these open questions. We enumerate them here for the sake of clarity:
\begin{enumerate}
\item [(a)] To determine which is the more appropriate temperature for the thermodynamic boundaries in Cosmology --- the Cai-Kim temperature \cite{Cai1} or the Hayward temperature \cite{Hayward1}.
\item [(b)] To determine whether the standard second law for the physical matter is ill-behaved so that we are forced to consider the GSL.
\item [(c)] To determine whether the GSL necessarily requires the synthetic assumption of local equilibrium.  
\item [(d)] The cosmic region bounded by the dynamical apparent horizon is a thermodynamic open system with the absolute Hubble flow crossing the horizon. What effect can this have on the entropy variation?
\item [(e)] To determine the consistency of the thermodynamic quantities with each other in gravitational thermodynamics.
\end{enumerate}
The paper has been organized as follows: The LDF model has been reviewed briefly in Section II. Section III concerns with the observational data analysis and the results of the analysis are presented in detail in Section IV. The GSL has been studied in Section V. Finally, a short discussion with conclusions can be found in Section VI. 
\section{The logotropic dark fluid model}\label{sec2}
In this section, we study the basic structure of the LDF Model. We assume a homogeneous and isotropic FLRW universe filled with a perfect fluid having energy density $\epsilon (t)$, rest mass density $\rho (t)$, and isotropic pressure $p(t)$. Further, we consider the Universe to be spatially flat as indicated by the anisotropy of the CMBR measurement \cite{flatk0}. Then, the Einstein's field equations yield the Friedmann and the acceleration equations\footnote{We humbly point out here that there are typos in the equations (2.1), (2.2), and (2.3) of Refs. \cite{lmc1,lmc3}, although the subsequent analyses are not affected by these typos.} given by \cite{Weinberg1}
\begin{equation}\label{eq1}
H^2 = \left(\frac{\dot{a}}{a}\right)^2 = \frac{8\pi G}{3}\epsilon 
\end{equation}
\begin{equation}\label{eq2}
\dot{H}+H^2 = \frac{\ddot{a}}{a} = -\frac{4\pi G}{3}\left(\epsilon +3\frac{p}{c^2}\right)
\end{equation}
respectively. In the above equation, $a(t)$ denotes the scale factor of the Universe, $H(t)=\left(\frac{\dot{a}}{a}\right)$ denotes the Hubble parameter and an overhead dot represents derivative with respect to the cosmic time $t$. Also, the constant parameters $G$ and $c$ represent the universal gravitational constant and the velocity of light respectively. Now, the energy conservation equation can be obtained as \cite{Weinberg1}
\begin{equation} \label{ECE}
\frac{d\epsilon}{dt}+3\left(\frac{\dot{a}}{a}\right)\left(\epsilon +\frac{p}{c^2}\right)=0.
\end{equation}
Among the above three equations (equations (\ref{eq1}), (\ref{eq2}) and (\ref{ECE})), only two are independent equations with three unknown parameters $H$, $\epsilon$ and $p$. So we still have freedom to choose one parameter to close the above system of
equations. For the present work, we assume that the Universe is filled with a single dark fluid satisfying an equation of state (EoS) \cite{lmc1,lmc2,lmc3,lmc4}
\begin{equation}\label{eqlogop}
p=A~\text{ln}\left(\frac{\rho}{\rho_{*}}\right),~~~A \geq 0
\end{equation}
which is known as the logotropic {\it equation of state} (EoS) and the fluid which obeys this EoS will be called the {\it logotropic dark fluid} (LDF). Here, $\rho$ is again the rest mass density, $A$ is the logotropic temperature (see Sec. 3 of Ref. \cite{lmc1}), and $\rho_{*}$ has been identified with the Planck density, $\rho_\text{P} = 5.16 \times 10^{99}\text{g~m}^{-3}$ (see Sec. 6 of Ref. \cite{lmc1}). The relation between the energy density $\epsilon$ and the rest mass density $\rho$ can be evaluated as \cite{lmc1}
\begin{eqnarray}
\epsilon &=& \rho c^2 + u(\rho) \nonumber \\
&=& \rho c^2 - A~\text{ln}\left(\frac{\rho}{\rho_\text{P}}\right) - A,
\end{eqnarray}
where $\rho c^2$ is the rest mass-energy and $u(\rho)=- A~\text{ln}\left(\frac{\rho}{\rho_\text{P}}\right) - A$, is the internal energy of the LDF respectively. Again, the pressure is related to the internal energy by the relation $p=-u-A$. Noting that a pressureless matter ($p=0$) gives $\rho = \rho_0 \left(\frac{a_0}{a}\right)^3$ from equation (\ref{ECE}), we have \cite{lmc1}
\begin{equation} \label{ep-1}
\epsilon = \rho_0 c^2 \left(\frac{a_0}{a}\right)^3 - A~\text{ln}\left(\frac{\rho_0}{\rho_\text{P}}\left(\frac{a_0}{a}\right)^3\right) - A,
\end{equation}
where the parameters with suffix '0' are their corresponding values at the present epoch. Chavanis \cite{lmc1} has shown that the first term in equation (\ref{ep-1}) mimics DM and the remaining terms mimics DE. We also observe that the early Universe ($a \rightarrow 0$, $\rho \rightarrow \infty$) was dominated by the rest mass-energy (DM), while the late Universe ($a \rightarrow \infty$, $\rho \rightarrow 0$) is dominated by the internal energy (DE). If we now introduce the dimensionless logotropic temperature $B=\frac{A}{\epsilon_{\Lambda}}$ and the normalized scale factor $R=\frac{a}{a_0}$, then equation (\ref{ep-1}) takes the equivalent form \cite{lmc1}
\begin{equation}\label{eqesl}
\frac{\epsilon}{\epsilon_0} = \frac{\Omega_{m0}}{R^3} + (1-\Omega_{m0})(1+3B~\text{ln}~R)
\end{equation}
where, $\epsilon_{0}=\frac{3H^{2}_{0}c^2}{8\pi G}$ is the present energy density of the Universe in which $H_0$ indicates the present value of the Hubble parameter. $\epsilon_{\Lambda}=(1-\Omega_{m0})\epsilon_0=\Omega_{\Lambda0}\epsilon_0$ is the present DE density, with $\Omega_{m0}$ and $\Omega_{\Lambda0}$ as the fractions of DM and DE at the present epoch, respectively.\\\\
Again, the pressure is related to the scale factor as \cite{lmc1}
\begin{equation}\label{eqp}
p=-\epsilon_0 (1-\Omega_{m0})(B+1+3B~\text{ln}~R).
\end{equation}
Finally using equations (\ref{eqesl}) and (\ref{eqp}), one can obtain the expression
for evolution equation of the EoS parameter $w$ for the LDF as 
\begin{equation} \label{ldfeos}
w=\frac{p}{\epsilon}=\frac{-(1-\Omega_{m0})(1+B+3B~\text{ln}~R)}{\frac{\Omega_{m0}}{R^3} + (1-\Omega_{m0})(1+3B~\text{ln}~R)}.
\end{equation}
The deceleration parameter also plays an important role in studying the evolutionary history of the Universe. It is defined as
\begin{equation}\label{eqqR}
q=-\frac{\ddot{a}}{aH^2} =-\frac{\dot{H}}{H^2}-1
\end{equation}
with the convention that the Universe will decelerate (${\ddot{a}}<0$) for $q>0$, while it will accelerate (${\ddot{a}}>0$) for $q<0$. Now, using equations (\ref{eq1}) and (\ref{eqesl}), the expression for the Hubble parameter can be obtained as 
\begin{equation}\label{eqhR}
H(a)= H_{0}\sqrt{\frac{\Omega_{m0}}{R^3} + (1-\Omega_{m0})(1+3B~\text{ln}~R)}.
\end{equation}
Then, using equations (\ref{eqqR}) and (\ref{eqhR}), the expression for $q$ is obtained as
\begin{equation}
q = \frac{\frac{\Omega_{m0}}{R^3} - (1-\Omega_{m0})(2+3B+6B~\text{ln}~R)}{2\left[\frac{\Omega_{m0}}{R^3} + (1-\Omega_{m0})(1+3B~\text{ln}~R)\right]}.
\end{equation}
\par In terms of redshift $z$, equation (\ref{eqhR}) can be written as
\begin{equation}\label{eqhz}
H(z)= H_{0}\sqrt{{\Omega_{m0}(1+z)^{3}} + (1-\Omega_{m0})(1-3B{\rm ln}(1+z))}
\end{equation}
where\footnote{We assume $a_0=1$, without any loss of generality.} $R=(1+z)^{-1}$. It is notable that for $B=0$, the logotropic model reduces to the standard $\Lambda$CDM model. It is important to mention here that the parameter $B$ depends on all the fundamental constants of physics and from now on, we shall regard $B$ as a fundamental constant (for details, see \cite{lmc1}). As a result, the present model only depends on two cosmological parameters $H_0$ and $\Omega_{m0}$, like the $\Lambda$CDM model. This interesting feature allows us to make a very accurate comparison between the two models in order to determine how close they are.\\  
\par Clearly, the cosmological characteristics of the LDF model given in equation (\ref{eqhz}) strongly depend on values of the parameters $H_{0}$ and $\Omega_{m0}$. In the next section, we have constrained these parameters ($H_{0}$ and $\Omega_{m0}$) using the latest observational data.
\section{Observational constraints on the model parameters}\label{data}
The Type Ia Supernova (SNIa) and cosmic chronometer (CC) datasets are very powerful in constraining various cosmological models. In this section, we shall fit the LDF model with the SNIa and CC datasets. For completeness, we have also described the datasets used in our analysis and the $\chi^2$ method used to analyze them.\\
\subsection{Cosmic chronometer (CC) data}
Being independent observational data, the structure of the expansion history of the universe can be well indicated by the $H(z)$ dataset \cite{sigmah}. From the observational point of view, the ages of the most massive and passively evolving galaxies, i.e. galaxies with old stellar populations and low star formation rates, will provide direct measurements of $H(z)$ at different redshifts \cite{jim2002}. These $H(z)$ measurements are independent of the Cepheid distance scale and do not depend on any specific cosmological model, although of course are subject to other systematic uncertainties. The galaxy differential age technique or CC approach was first introduced in  \cite{jim2002} to measure $H(z)$. It uses the relative ages of the most massive and passively evolving galaxies to measure $\frac{dz}{dt}$. The Hubble parameter depending on the differential ages as a function of redshift $z$ can be written in the form of
\begin{equation}
H(z)=-\frac{1}{(1+z)}\frac{dz}{dt} 
\end{equation}
It is evident from the above equation that $H(z)$ can be obtained directly if $\frac{dz}{dt}$ is known. For a given pair of ensembles of passively-evolving galaxies at two different redshifts it is possible to deduce the derivative $\frac{dz}{dt}$ using the spectroscopic dating techniques \cite{simon2005}. As discussed in \cite{simon2005}, the measurements of the age difference ($\bigtriangleup t$), between two passively-evolving galaxies that formed at the same time but are separated by a small
redshift interval ($\bigtriangleup z$), one can deduce $\frac{dz}{dt}$, from the ratio $\frac{\bigtriangleup z}{\bigtriangleup t}$. Therefore, CC approach allow us to obtain direct information about the Hubble parameter at various redshifts, contrary to other probes which do not directly measure Hubble parameter, but integrated quantities as e.g. luminosity distances. In this work, we have used the latest observational $H(z)$ dataset obtained through the CC approach, consisting of 31 data points in the redshift range, $0< z < 2$ \cite{hzdataMore,hzdataMeng,hzdatarefcao,hzdatarefzhang,hzdatarefstern} and the corresponding $H(z)$ values are given in the Table I of \cite{31hztab}. Note that here we do not make use of dataset on $H(z)$ obtained from the measurement of baryon acoustic oscillations in order to avoid dealing with their cosmological model dependence. For this dataset, the $\chi^{2}$ function is defined as 
\begin{equation}\label{eqchi2h}
\chi^2_{CC} = \sum^{31}_{i=1}\frac{[{H}^{obs}(z_{i}) - {H}^{th}(z_{i},\theta_m)]^2}{\sigma^2_{H}(z_{i})} 
\end{equation}
where $\sigma_{H}(z_{i})$ represents the error associated with the $i^{th}$ data point and $\theta_m$ denotes the model parameters. Hereafter, the subscript ``obs" refers to observational quantities and subscript ``th" refers to the corresponding theoretical ones. 
\subsection{Supernovae type Ia (SNIa) data}
Next, we use 1048 Supernovae data points from the compilation of Pantheon sample available in \cite{pantsnia}, in
the redshift range $0.01 < z < 2.3$. The $\chi^2$ for this dataset is given by
\begin{equation}
\chi^{2}_{SN}(\theta_{m})= \sum^{1048}_{i,j=1}\vartriangle \mu_{i}. (C^{-1}_{SN})_{ij}.\vartriangle \mu_{j}
\end{equation}
where $\vartriangle \mu_{i} = \mu_{th} (z_{i}, \theta_{m}) - \mu_{obs} (z_{i})$, $\theta_{m}$ , $C_{SN}$ are respectively the discrepancy in distance modulus 
between theory and observations, model parameters to be fitted, and the covariance matrix \cite{pantsnia}.\\

 Then, we use the maximum likelihood method and take the likelihood function as
\begin{equation}
{\cal L} ={\rm e}^{-\frac{\chi^2_{t}}{2}}
\end{equation}
where, $\chi^2_{t}=\chi^2_{CC}+\chi^2_{SN}$. It should be noted that the best-fit parameter values (say, $\theta^{*}_{m}$) are those that maximize the likelihood function (or minimize the $\chi^2$ function )
\begin{equation}
{\cal L}(\theta^{*}_{m})={\rm e}^{-\frac{\chi^2_{t}(\theta^{*}_{m})}{2}}
\end{equation}
We can now plot the contours for different confidence levels. The confidence levels $1\sigma(68.3\%)$ and $2\sigma(95.4\%)$ are taken proportional to $\bigtriangleup \chi^{2}=2.3$ and $6.17$ respectively, where $\bigtriangleup \chi^{2}= \chi^{2}_{t}(\theta_m)- \chi^{2}_{t}(\theta^{*}_{m})$ and $\chi^{2}_{min}$ is the minimum value of $\chi^2_{t}$. The fit is good and the data are well consistent with the LDF model, if 
\begin{equation}
{\chi}^{2}_{r}=\frac{\chi^{2}_{min}}{N_{dof}}\leq 1
\end{equation}
where, $N_{dof}$ denotes the degree of freedom and it is defined as the difference between all observational data points and the number of free parameters.
In what follows, we describe the main observational consequences for the LDF model.
\section{Results of the Data Analysis}\label{result} 
\begin{figure*}
\resizebox{6cm}{!}{\rotatebox{0}{\includegraphics{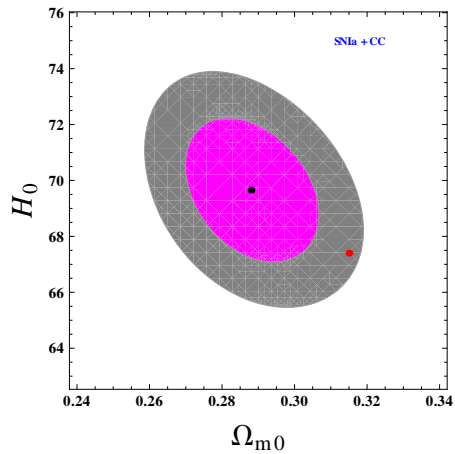}}}
\caption{This figure shows the $1\sigma$ (magenta region) and $2\sigma$ (gray region) confidence contours in the $\Omega_{m0}$-$H_{0}$ plane using the SNIa+CC dataset. In this plot, the black dot represents the best-fit values of the pair ($\Omega_{m0}$, $H_{0}$) for the present model. Also, the red point represents the best-fit values of the parameters $\Omega_{m0}=0.315$ and $H_{0}=67.4$, obtained by  
the Planck analysis \cite{gh01} assuming the base-$\Lambda$CDM cosmology.}
\label{figc}
\end{figure*}

\begin{figure*}
\resizebox{6cm}{!}{\rotatebox{0}{\includegraphics{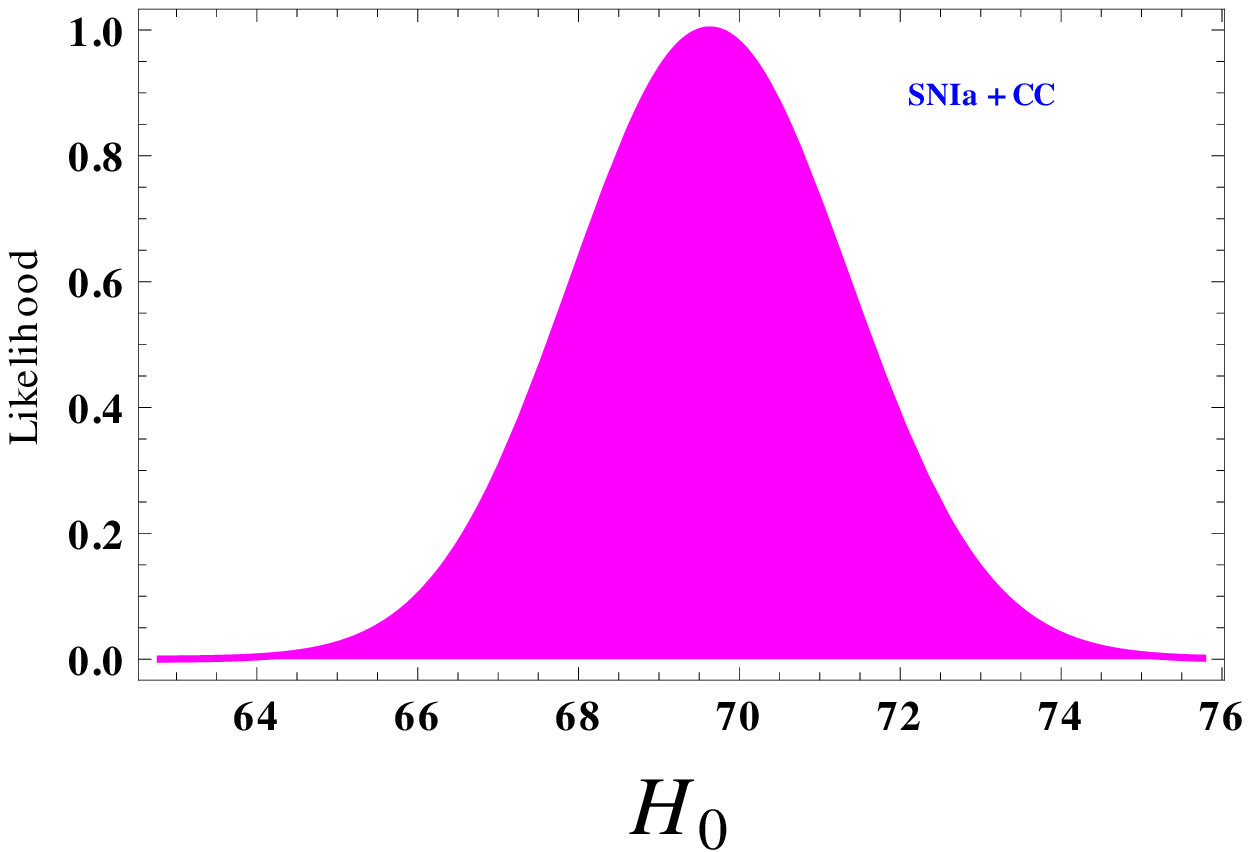}}}\hspace{5mm}
\resizebox{6cm}{!}{\rotatebox{0}{\includegraphics{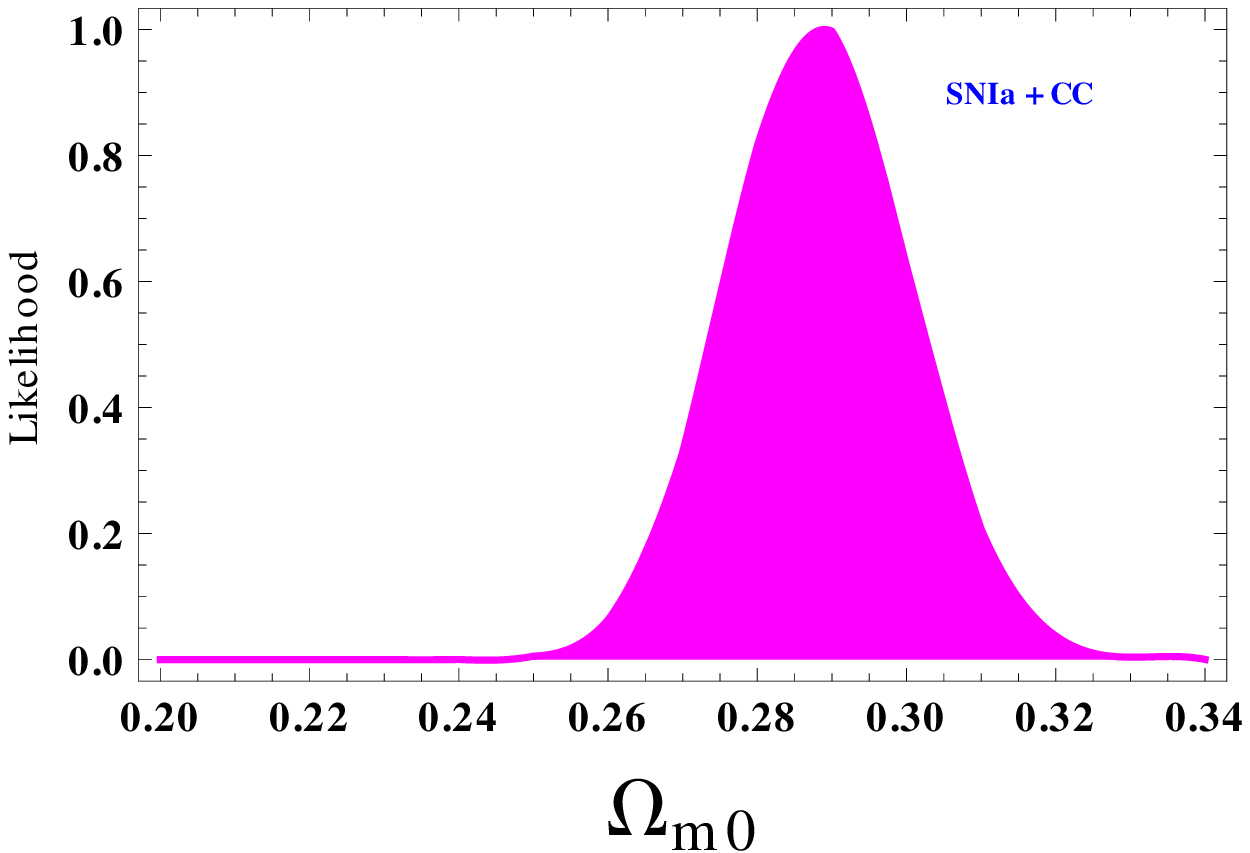}}}
\caption{Left panel shows the marginalized likelihood function vs. $H_0$ and the right panel shows the marginalized likelihood function vs. $\Omega_{m0}$ for the present model.}
\label{figl}
\end{figure*}

\begin{figure*}
\resizebox{6cm}{!}{\rotatebox{0}{\includegraphics{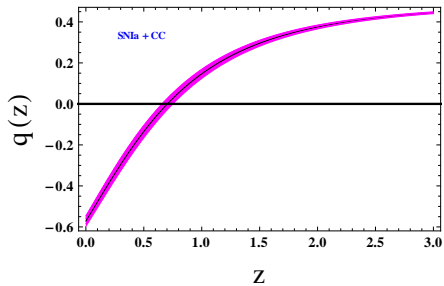}}}
\caption{The dynamical evolution of $q$ as a function of $z$ is shown in $1\sigma$ confidence region. Here, the central dark line denotes the best-fit curve resulting from our joint analysis, while the horizontal line denotes $q(z)=0$. The intersection  of  the  best-fit curve  with  the  horizontal  line  corresponds  to the point at which the Universe starts accelerating.}
\label{figq}
\end{figure*}

\begin{figure*}
\resizebox{6cm}{!}{\rotatebox{0}{\includegraphics{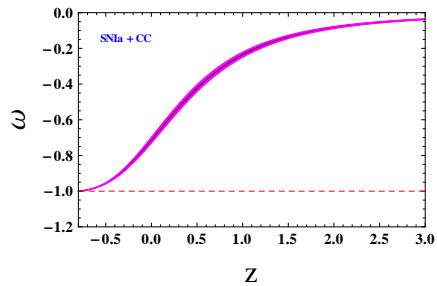}}}
\caption{The evolution of the logotropic equation of state parameter $w$ against $z$ is shown in $1\sigma$ confidence region for the present model. Here, the central dark line denotes the best-fit curve, while the horizontal line (red dashed) represents the $\Lambda$CDM $(w_{\Lambda}=-1)$ model.}
\label{figw}
\end{figure*}

\begin{figure*}
\resizebox{6cm}{!}{\rotatebox{0}{\includegraphics{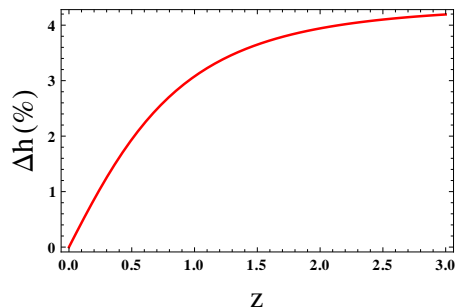}}}
\caption{This figure shows the percentage deviation in the normalized Hubble parameter $h$ as function of $z$ compared to the $\Lambda$CDM model. In this plot, we have taken $\Omega_{m0}=0.315$ \cite{gh02} for the $\Lambda$CDM model.}
\label{figdhp}
\end{figure*}

In this section, we have discussed the results obtained from the $\chi^{2}$ analysis method (as described in the previous section). We have obtained the constraints on the model parameters $\Omega_{m0}$ and $H_{0}$ by using the latest Pantheon SNIa+CC dataset. It is important to mention here that for the present analysis and in the all figures (Fig. \ref{figc}-\ref{figdhp}), we have considered the value $B=3.53 \times 10^{-3}$, as predicted by the theory in Refs.\cite{lmc1,lmc2}. The $1\sigma$ and $2\sigma$ contours in $\Omega_{m0}-H_{0}$ plane for the LDF model is shown in figure \ref{figc}. The mean values for the model parameters are obtained as $\Omega_{m0}=0.288\pm0.012$ and $H_{0}=69.652\pm 1.698~{\rm km~s^{-1}~Mpc^{-1}}$ (with $\chi^{2}_{r}=0.975$). It has been observed that the value of $\Omega_{m0}$ obtained in this work is slightly lower than the value obtained by the Planck analysis \cite{gh02}, which puts the limit on $\Omega_{m0}$ as $\Omega_{m0}=0.315\pm 0.017$ with $1\sigma$ errors \cite{gh02}. We have also found from figure \ref{figc} that the best estimate values of the parameters $\Omega_{m0}$ and $H_0$ (as shown by the red dot) from the Planck analysis \cite{gh01,gh02}, are found to be well within the $2\sigma$ confidence contour. Interestingly, it has been found that the mean value of the parameter $H_0$ obtained in the present analysis is almost same with the value $H_{0}=70.5^{+0.5}_{-0.5}~{\rm km~s^{-1}~Mpc^{-1}}$, obtained by the Lin et al. \cite{linh0}, using the Pantheon compilation of type Ia supernovae and the non-parametric method. In a relevant work, Capozziello et al. \cite{lmc7} studied a new class of single dark fluid model and obtained $H_{0}=65.67^{+1.75}_{-1.78}~{\rm km~s^{-1}~Mpc^{-1}}$ using the JLA SNIa+CC+BAO dataset. Thus, the present work provides better constraint on $H_{0}$ as compared to the results obtained in Ref. \cite{lmc7}, which has the LDF model as a particular case. It deserves to mention here that the Pantheon sample 
is the largest spectroscopically confirmed SNIa sample to date and comparing to the {\it joint light curve} (JLA) SNIa data, the Pantheon SNIa data can give tighter dark energy constraints (for details, one can look into Ref. \cite{panjla}). Therefore, it is reasonable to expect the improvement in our observational analysis due to the use of the high quality Pantheon SNIa data instead of the JLA SNIa sample. On the other hand, it is well known that there is more than $3\sigma$ tension between the values of $H_0$ measured from the global CMB radiation ($H_{0}=67.4^{+0.5}_{-0.5}~{\rm km~s^{-1}~Mpc^{-1}}$ \cite{gh01}, $H_{0}=67.3^{+1.2}_{-1.2}~{\rm km~s^{-1}~Mpc^{-1}}$ \cite{gh02}) and that from the local distance ladders ($H_{0}=74.03^{+1.42}_{-1.42}~{\rm km~s^{-1}~Mpc^{-1}}$ \cite{lh01}). In fact, there are many attempts to alleviate the $H_0$ tension problem and some of the recent important works in this topic can be found in Refs. \cite{h0t1,h0t2,h0t3,h0t4,h0t5,h0t6}. The most interesting result of our anlysis is that the inferred Hubble constant is approximately the mean value of the global and local measurements of $H_0$, thus may alleviate the tension between the global and local measurements. Also, the marginalized likelihoods of individual parameters are shown in figure \ref{figl}. It is clear from the likelihood plots that the likelihood functions are well fitted to a Gaussian distribution function for the combined SNIa+CC dataset. \\
\par The plot of the deceleration parameter $q(z)$, as given in figure \ref{figq}, clearly shows that the LDF model successfully generates late time cosmic acceleration ($q<0$) along with a decelerated ($q>0$) expansion phase in the past. This is essential for the structure formation of the Universe. It is observed that $q(z)$ shows a signature flip at the transition redshift $z_{t}=0.706\pm 0.048$ (within $1\sigma$ error). From figure \ref{figq}, the present value of $q$ is found to be $q_{0}=-0.572\pm 0.026$. These  results  are  in  good agreement  with  the  recent  estimate found in Refs. \cite{zt1,zt2,zt3,zt4,zt5,zt6,zt7,zt8,zt9,zt10}. Furthermore, the functional behavior of the equation of state parameter $w$ is displayed on figure \ref{figw}. From this figure, one can clearly observe that for the best-fit model, the value of $w$ was close to zero at the high redshifts,  at the current epoch (i.e., $z=0$) it is close to $-0.715\pm0.017$ (within $1\sigma$ error) and settles to a value $-1$ in future. 
Thus, it is also evident that there is no future singularity in this model. These scenarios also agree very well with the results obtained in Refs. \cite{acc6,fsl,fslaam}.  For a comprehensive analysis, in figure \ref{figdhp}, we have also plotted the percentage deviation in the normalized Hubble parameter ${\left(\triangle h (\%)=\triangle {\left(\frac{H(z)}{H_0}\right)}=\frac{h(z)-h_{\Lambda CDM}(z)}{h_{\Lambda CDM}(z)}~\times~100\right)}$ for the above model as compared to a $\Lambda$CDM model, and the corresponding deviation is found to be $0.83\%$ at $z\sim~0.2$, $1.9\%$ at $z\sim~0.5$ and $3.65\%$ at $z\sim~1.5$. Again, we have also found that the two models are indistinguishable at present. Therefore, these deviations in the present model also need further attention because the dark components (DE and DM) are two manifestations of the same dark fluid. As a result, it may solve or at least alleviate the cosmological coincidence problem.
\section{Generalized Second Law in the Logotropic Model}\label{gsl}
This section deals with a study of the {\it generalized second law} (GSL) of thermodynamics at the dynamical apparent horizon in the Logotropic model. We consider the actual expression (i.e., the non-truncated version) of the Hawking temperature \cite{Hawking1} at the apparent horizon\footnote{Recall that in a flat FLRW universe, the radius $R_{A_{\text{L}}}$ of the apparent horizon is simply $R_{A_{\text{L}}} = \frac{1}{H}$.}
\begin{equation}
T_{A_{\text{L}}}=\frac{1}{2\pi R_{A_{\text{L}}}}\left(1-\frac{\dot{R}_{A_{\text{L}}}}{2}\right),
\end{equation} 
where $R_{A_{\text{L}}}$ is the proper radius of the apparent horizon in the Logotropic model. Although, the truncated expression
\begin{equation}
T_{A_{\text{L}}}=\frac{1}{2\pi R_{A_{\text{L}}}}
\end{equation}
is generally used in gravitational thermodynamics but Bi{\'e}truy and Helou \cite{Binetruy1,Helou1} has put forward several strong arguments against using this type of formalism. Also, the use of the former expression has led to some promising results recently \cite{Saha1}. We shall consider two types
of entropies on the dynamical apparent horizon, the most widely used Bekenstein entropy \cite{Bekenstein1} and the recently proposed Viaggiu entropy \cite{Viaggiu1,Viaggiu2}. It is interesting to note that although the Viaggiu entropy is simply a correction to the Bekenstein entropy due to the dynamic nature of the Universe, yet it has yielded a very nice result for a constant EoS paramter \cite{Saha1} which is in striking contrast with that obtained with the Bekenstein entropy \cite{Wang1,Saha2,Saha3}. With this thermodynamic analysis, we aim to garner support for the choice of the value of the free parameter $B$ in the Logotropic model. Chavanis has already obtained the exact value $3.53 \times 10^{-3}$ for $B$ from the measurement of surface density of DM halos \cite{lmc1} but here we employ ourselves in finding a range of values for $B$ purely by thermodynamic means. If we succeed, the Logotropic model will be put in a much stronger footing. This is due to the well-known fact that there exists an intimate connection between thermodynamics and General Relativity \cite{Jacobson1,Padmanabhan1,Padmanabhan2,Cai1,Gong1}.\\ 

Let us first note that the Bekenstein entropy on the horizon $R_{A_{\text{L}}}$ has the expression \cite{Bekenstein1}
\begin{eqnarray} \label{Bek-ent}
S_{A_{\text{L}}}^{B} &=& \left(\frac{c^3}{G\hbar}\right)\frac{A_{A_{\text{L}}}}{4} \nonumber \\
&=& \pi R_{A_{\text{L}}}^2,
\end{eqnarray}
where\footnote{Henceforth, in this section, we assume gravitational units $G=c=\hbar=\kappa_B= 1$.} $A_{A_{\text{L}}}=4\pi R_{A_{\text{L}}}^2$ is the proper area bounded by the apparent horizon, while the Viaggiu entropy on the horizon $R_{A_{\text{L}}}$ is expressed as \cite{Viaggiu1,Viaggiu2,Saha1}
\begin{eqnarray} \label{Viag-ent}
S_{A_{\text{L}}}^{V} &=& \left(\frac{1}{4L_{p}^{2}}\right) A_{A_{\text{L}}} + \left(\frac{3\kappa _B}{2cL_{p}^{2}}\right) V_{A_{\text{L}}} H \nonumber \\
&=& \pi R_{A_{\text{L}}}^2 + 2\pi R_{A_{\text{L}}}^2,
\end{eqnarray}
where $L_{p} = \sqrt{\frac{G\hbar}{c^3}}$ is the Planck length and $V_{A_{\text{L}}}=\frac{4}{3}\pi R_{A_{\text{L}}}^3$ is the volume bounded by the apparent horizon. The time-derivative of entropy of the fluid inside the horizon, $\dot{S}_{fA_{\text{L}}}$, is evaluated by using the Clausius relation
\begin{equation} \label{fl-ent}
T_{fA_{\text{L}}}dS_{fA_{\text{L}}} = dU + pdV_{A_{\text{L}}},
\end{equation}
where $T_{fA_{\text{L}}}$ and $S_{fA_{\text{L}}}$ are, respectively, the temperature and the entropy of the fluid, while $U = \frac{4}{3}\pi R_{A_{\text{L}}}^3 \epsilon$ is the internal energy of the fluid, evaluated at the dynamical apparent horizon.\\

Using equations (\ref{Bek-ent}) and (\ref{fl-ent}), we arrive at the total entropy\footnote{In calculating the total entropy, we assume that the temperature of the horizon and that of the fluid inside are equal, in accordance with the pioneering work by Mimoso and Pav{\'o}n \cite{Mimoso1}.} (for the case with Bekenstein entropy and non-truncated Hawking temperature; equation (19) of Ref. \cite{Saha1})
\begin{equation}
\dot{S}_{TA_{\text{L}}}^{B} = 18\pi R_{A_{\text{L}}} \frac{(1+w)^2}{(1-3w)},
\end{equation}
which shows that the GSL is true for $w \leq \frac{1}{3}$. On the other hand, using equations (\ref{Viag-ent}) and (\ref{fl-ent}), the expression for total entropy becomes (for the case with Viaggiu entropy and non-truncated Hawking temperature; equation (21) of Ref. \cite{Saha1})
\begin{equation}
\dot{S}_{TA_{\text{L}}}^{V} = 6\pi R_{A_{\text{L}}} \frac{(1+w)(8+3w)}{(1-3w)},
\end{equation}
from which it has been established \cite{Saha1} that the GSL holds only for $-1 \leq w \leq \frac{1}{3}$.\\

We carefully observe here that the upper bounds in both the cases are the same but the Viaggiu entropy, in addition, forces a lower bound on the value of the EoS parameter. These two inequalities are actually equivalent to a single inequality $-1 \leq w \leq \frac{1}{3}$. Now, replacing $w$ by the LDF EoS given in equation (\ref{ldfeos}), we arrive at
$$-1 \leq \frac{-(1-\Omega_{m0})(B+1+3B~\text{ln}~R)}{\frac{\Omega_{m0}}{R^3}+(1-\Omega_{m0})(1+3B~\text{ln}~R)} \leq \frac{1}{3}.$$
After doing some algebra with inequalities, we can restrict the parameter $B$:
\begin{equation}\label{eqtcB}
\frac{-4+3\Omega_{m0}}{3(1-\Omega_{m0})} \leq B \leq \frac{\Omega_{m0}}{1-\Omega_{m0}}
\end{equation}
at the present epoch, $R=1$. For the best-fit model, i.e., $\Omega_{m0}=0.288$, we finally obtain (using equation (\ref{eqtcB})) 
\begin{equation}\label{eqthbbfB}
-1.468 \leq B \leq 0.404.
\end{equation}
Since we have considered $B=\frac{A}{\epsilon_\Lambda}$ and $A \geq 0$, we must have $B \geq 0$. This implies that $0 \leq B \leq 0.404$. Therefore, the value $B=3.53 \times 10^{-3}$ obtained in Ref. \cite{lmc1} from  galactic observations, is consistent with thermodynamics. We also observe that the upper bound on $B$, as given in equation (\ref{eqthbbfB}), is slightly higher than the corresponding bounds on $B$ ($0 \leq B \leq 0.09425$, $0 \leq B \leq 0.0262$ and $0 \leq B \leq 0.0379$), as obtained in Ref. \cite{lmc1} from the galactic observations and from the measurements of the CMB shift parameter respectively. It is also interesting to see that we have obtained a range of allowable values of $B$ purely by thermodynamic means. 
\section{Conclusions}\label{conclusion}
We have considered a spatially flat, homogeneous and isotropic FLRW Universe filled with a single dark fluid, whose pressure evolves through a logarithmic equation of state, as given in equation (\ref{eqlogop}). The theoretical motivations and interesting features of this unified model have already been discussed in details in sections \ref{secintro} $\&$ \ref{sec2}. We have then constrained the free parameters of the model by $\chi^{2}$ minimization technique using the Pantheon SNIa+CC dataset. In particular, we have obtained $\Omega_{m0}=0.288\pm 0.012$ and $H_{0}=69.652\pm 1.698~{\rm km~s^{-1}~Mpc^{-1}}$, which is in agreement with the recent estimate obtained $H_{0}=70.5^{+0.5}_{-0.5}~{\rm km~s^{-1}~Mpc^{-1}}$ in Ref. \cite{linh0}. Additionally, we have also found that the present analysis provides better constraint on $H_{0}$ as compared to the results reported in Ref. \cite{lmc7}, which has the LDF model as a particular case. As mentioned in section \ref{result}, our detailed study shows that the deduced Hubble constant $H_0$ is approximately the mean value of the global and local measurements of $H_0$, and thus may alleviate the tension between these measurements. We have also investigated the epoch of the DE dominance that drives the accelerated expansion of the Universe. It has been found that the values of the transition redshift (from decelerated to accelerated expansion) obtained within $1\sigma$ confidence level, are in good agreement with the previous results as reported in Refs. \cite{zt1,zt2,zt3,zt4,zt5,zt6,zt7,zt8,zt9,zt10}. Additionally, we have also compared the Logotropic and $\Lambda$CDM models in order to determine quantitatively how much they differ. We have found that there is no significant difference between the Logotropic and $\Lambda$CDM models at the present epoch, but the difference between these models is evident at high redshifts (see figure \ref{figdhp}). This may provide a possible solution to a number of cosmological problems. \\
\par Furthermore, we have undertaken a thermodynamic study of the Logotropic model at the dynamical apparent horizon by considering Bekenstein entropy \cite{Bekenstein1} and Viaggiu entropy \cite{Viaggiu1,Viaggiu2}. We have restricted our study to the generalized second law of thermodynamics only. It has been found that for the case of Bekenstein entropy, the GSL of thermodynamics holds for $w \leq \frac{1}{3}$, while for the case of Viaggiu entropy, the GSL holds only for  $-1 \leq w \leq \frac{1}{3}$. As mentioned earlier, the model studied in this work depends on the parameter $B$ (dimensionless logotropic temperature) in such a way that for $B = 0$, the $\Lambda$CDM model is recovered. Using the best-fit value of $\Omega_{m0}$, we have also obtained a thermodynamically allowable range for the parameter $B$, $0 \leq B \leq 0.404$. This result is interesting from both observational as well as cosmological points of view. It is important to note that these bounds support our earlier choice of its value, $B=3.53 \times 10^{-3}$ for which we have plotted the graphs in section \ref{result}. We reiterate here that this particular value was obtained by P.H. Chavanis \cite{lmc1} from the surface density of DM halos.\\ \\
According to the aforementioned results, we note that the present model is reliable for further study and is compatible with the latest SNIa and CC observational dataset. Finally, we conclude that our model seems to represent a viable alternative to the $\Lambda$CDM model.
\section{ACKNOWLEDGMENTS}
We are grateful to the anonymous referee for constructive criticisms and kind suggestions which have helped us to improve this work significantly.


\begin{thebibliography}{99}
\bibitem{acc1} S. Perlmutter et al., Astrophys. J. {\bf 517}, 565 (1999).
\bibitem{acc2}A. G. Riess et al., Astron. J. {\bf 116}, 1009 (1998).
\bibitem{acc3}M. Tegmark et al., Phys. Rev. D {\bf 69}, 103501 (2004). 
\bibitem{acc4}U. Seljak et al., Phys. Rev. D {\bf 71}, 103515 (2005). 
\bibitem{acc5}D. J. Eisenstein et al., Astrophys. J. {\bf 633}, 560 (2005).
\bibitem{acc6}E. Komatsu et al., Astrophys. J. Suppl. {\bf 192}, 18 (2011).
\bibitem{acc7}G. Hinshaw et al., Astrophys. J. Suppl. {\bf 208}, 19 (2013).
\bibitem{acc8}P. A. R. Ade et al., Phys. Rev. Lett. {\bf 112}, 241101 (2014).
\bibitem{acc9}P. A. R. Ade et al., Phys. Rev. Lett. {\bf 114}, 101301 (2015).
\bibitem{acc10}P. A. R. Ade et al., Astron. Astrophys. {\bf 594}, A13 (2016).
\bibitem{de1}E. J. Copeland, M. Sami and S. Tsujikawa, Int. J. Mod. Phys. D {\bf 15}, 1753 (2006).
\bibitem{de2}P. J. E. Peebles and B. Ratra, Rev. Mod. Phys. {\bf 75}, 559 (2003).
\bibitem{de3}V. Sahni and A. A. Starobinsky, Int. J. Mod. Phys. D {\bf 9}, 373 (2000).
\bibitem{ccp1}S. Weinberg, Rev. Mod. Phys. {\bf 61}, 1 (1989).
\bibitem{ccp2}P.J. Steinhardt et al., Phys. Rev. Lett. {\bf 59}, 123504 (1999).
\bibitem{lmc1}P.H. Chavanis, Eur. Phys. J. Plus {\bf 130}, 130 (2015).
\bibitem{lmc2}P.H. Chavanis, Phys. Lett. B {\bf 758}, 59 (2016).
\bibitem{lmc3} P.H. Chavanis and S. Kumar, JCAP {\bf 1705}, 018 (2017).
\bibitem{lmc4} P.H. Chavanis, Phys. Dark Univ. {\bf 24}, 100271 (2018).
\bibitem{lmc5}V. M. C. Ferreira and P. P. Avelino, Phys.Lett. B {\bf 770}, 213 (2017).
\bibitem{lmc6}K. Boshkayev, R. D'Agostino and O. Luongo, Eur.Phys.J. C {\bf 79}, 332 (2019).
\bibitem{lmc7}S. Capozziello, R. D'Agostino and O. Luongo, Phys. Dark Univ. {\bf 20}, 1 (2018).
\bibitem{Burkert1} A. Burkert, Astrophys. J. {\bf 447}, L25 (1995).
\bibitem{Navarro1} J.F. Navarro, C.S. Frank, and S.D.M. White, Astrophys. J. {\bf 462}, 563 (1996).
\bibitem{Donato1} F. Donato et al., Mon. Not. Roy. Astron. Soc. {\bf 397}, 1169 (2009).
\bibitem{Strigari1} L. E. Strigari et al., Nature {\bf 454}, 1096 (2008).
\bibitem{Tully1} R.B. Tully and J.R. Fisher, Astron. Astrophys. {\bf 54}, 661 (1977).
\bibitem{udem1}A. Kamenshchik, U. Moschella and V. Pasquier , Phys. Lett. B {\bf 511}, 265 (2001).
\bibitem{udem2}M. C. Bento, O. Bertolami and A.A.Sen , Phys. Rev. D {\bf 66}, 043507 (2002).
\bibitem{udem3} M.C. Bento, O. Bertolami and A. A. Sen, Phys. Rev. D {\bf 70}, 083519 (2004).
\bibitem{udem4} V. Gorini, A. Kamenshchik, U. Moschella, Phys. Rev. D 67, 063509 (2003).
\bibitem{udem5}P.H. Chavanis, Eur. Phys. J. Plus {\bf 129}, 38 (2014).
\bibitem{Bekenstein1} J.D. Bekenstein, Phys. Rev. D {\bf 7}, 2333 (1973).
\bibitem{Viaggiu1} S. Viaggiu, Mod. Phys. Lett. A {\bf 29}, 1450091 (2014).
\bibitem{Viaggiu2} S. Viaggiu, Gen. Relativ. Gravit. {\bf 47}, 86 (2015).
\bibitem{Tian1} D. W. Tian and I. Booth, Phys. Rev. D {\bf 92}, 024001 (2015).
\bibitem{Cai1} R. G. Cai and S. P. Kim, JHEP {\bf 0502}, 050 (2005).
\bibitem{Hayward1} S. A. Hayward, Class. Quantum Grav. {\bf 15}, 3147 (1998).
\bibitem{flatk0}P. de Bernardis et al., Nature, {\bf 400}, 955, (2000).
\bibitem{Weinberg1} S. Weinberg, {\it Gravitation and Cosmology} (John Wiley, USA, 2002).
\bibitem{sigmah}M. Seikel, S. Yahya, R. Maartens, C. Clarkson, Phys. Rev. D {\bf 86}, 083001 (2012).
\bibitem{jim2002}R. Jimenez, A. Loeb,  Astrophys. J., {\bf 573}, 37 (2002).
\bibitem{simon2005}J. Simon, L. Verde and R. Jimenez,  Phys. Rev. D., {\bf 71}, 123001 (2005).
\bibitem{hzdataMore}M. Moresco et al., JCAP, {\bf 05}, 014 (2016).
\bibitem{hzdataMeng}X.-L. Meng et al., arXiv:1507.02517 (2015).
\bibitem{hzdatarefcao}S.-L. Cao et al., Eur. Phys. J. C, {\bf 78}, 313 (2018).
\bibitem{hzdatarefzhang}C. Zhang et al., A \& A, {\bf 14}, 1221 (2014).
\bibitem{hzdatarefstern}D. Stern et al., J. Cosmol. Astropart. Phys., {\bf 2}, 8 (2010).
\bibitem{31hztab}J. Magana et al., MNRAS, {\bf 476}, 1036 (2018).
\bibitem{pantsnia}D.M. Scolnic et al., APJ {\bf 859}, 101 (2018).
\bibitem{gh02} P. A. R. Ade et al. [Planck Collaboration], A$\&$A, {\bf 571}, A16 (2014).
\bibitem{gh01}N. Aghanim et al. [Planck Collaboration], arXiv:1807.06209 [astro-ph.CO] (2019).
\bibitem{linh0}H. N. Lin, X. Li and L. Tang, Chin. Phys. C {\bf 43}, 075101 (2019).
\bibitem{panjla}S. Wang and X. Luo, arXiv: 1912.11879 (2019). 
\bibitem{lh01} A. G. Riess, S. Casertano, W. Yuan, L. M. Macri
and D. Scolnic, Astrophys. J. {\bf 876}, 85 (2019).
\bibitem{h0t1} E. Di Valentino, A. Melchiorri and J. Silk, JCAP {\bf 2001}, 013 (2020).
\bibitem{h0t2}E. Di Valentino, A. Melchiorri, O. Mena and S. Vagnozzi, Phys. Rev. D {\bf 101}, 063502 (2020). 
\bibitem{h0t3}J. L. Bernal, L. Verde and A. G. Riess, JCAP {\bf 1610}, 019 (2016).
\bibitem{h0t4}R. Jimenez, A. Cimatti, L. Verde, M. Moresco, B. Wandelt, JCAP {\bf 03}, 043 (2019).
\bibitem{h0t5}K. C. Wong, S. H. Suyu, G. C.-F. Chen and E. Komatsu et al.,  	arXiv:1907.04869 (2020) [Accepted in MNRAS, DOI: 10.1093/mnras/stz3094]. 
\bibitem{h0t6}I. Jee, S. Suyu, E. Komatsu, C. D. Fassnacht, S. Hilbert, L. V. E. Koopmans, Science {\bf 365}, 6458 (2019).
\bibitem{zt1}D. Rapetti, S. W. Allen, M. A. Amin and R. D. Blandford, Mon. Not. Roy. Astron. Soc.,{\bf 375}, 1510 (2007).
\bibitem{zt2}R. Nair et al., J. Cosmol. Astropart. Phys., {\bf 01}, 018 (2012).
\bibitem{zt3}O. Farooq, B. Ratra, Astrophys. J., {\bf 766}, L7 (2013).
\bibitem{zt4}J. Magana et al., J. Cosmol. Astropart. Phys., {\bf 10}, 017 (2014).
\bibitem{zt5}S. Capozziello, O. Farooq, O. Luongo, B. Ratra, Phys. Rev. D., {\bf 90}, 044016 (2014).
\bibitem{zt6}A. A. Mamon, S. Das, Int. J. Mod. Phys. D., {\bf 25}, 1650032 (2016).
\bibitem{zt7}A. A. Mamon, S. Das, Eur. Phys. J. C, {\bf 77}, 495 (2017).
\bibitem{zt8}O. Farooq, F. R. Madiyar, S. Crandall, B. Ratra, {\bf ApJ}, 835, 26 (2017).
\bibitem{zt9}A. A. Mamon, K. Bamba, S. Das, Eur. Phys. J. C, {\bf 77}, 29 (2017). 
\bibitem{zt10} A. A. Mamon and K. Bamba, Eur.Phys.J. C {\bf 78}, 862 (2018).
\bibitem{fsl} M. Sullivan et al., Astrophys. J. {\bf 737}, 102 (2011).
\bibitem{fslaam} S. Das and A. A. Mamon, Astrophys Space Sci. {\bf 355}, 371 (2015).
\bibitem{Hawking1} S. W. Hawking, Commun. Math. Phys. {\bf 43}, 199 (1975).
\bibitem{Binetruy1} P. Binetruy and A. Helou, Class. Quantum Grav. {\bf 32}, 205006 (2015).
\bibitem{Helou1} A. Helou, arXiv:1502.04235 [gr-qc].
\bibitem{Saha1} S. Saha, Int. J. Mod. Phys. A {\bf 34}, 1950193 (2019).
\bibitem{Wang1} B. Wang, Y. Gong, and E. Abdalla, Phys. Rev. D {\bf 74}, 083520 (2006).
\bibitem{Saha2} S. Saha and S. Chakraborty, Phys. Lett. B {\bf 717}, 319 (2012).
\bibitem{Saha3} S. Saha and S. Chakraborty, Phys. Rev. D {\bf 89}, 043512 (2014).
\bibitem{Jacobson1} T. Jacobson, Phys. Rev. Lett. {\bf 75}, 1260 (1995).
\bibitem{Padmanabhan1} T. Padmanabhan, Class. Quantum Grav. {\bf 19}, 5387 (2002).
\bibitem{Padmanabhan2} T. Padmanabhan, Phys. Rep. {\bf 406}, 49 (2005).
\bibitem{Gong1} Y. Gong and A. Wang, Phys. Rev. Lett. {\bf 99}, 211301 (2007).
\bibitem{Mimoso1} J. P. Mimoso and D. Pav{\'o}n, Phys. Rev. D {\bf 94}, 103507 (2016).
\end{thebibliography}
\end{document}